\documentstyle[preprint,aps]{revtex}
\begin{document}
\title{ \bf REMARKS ON THE COSMOLOGICAL CONSTANT AND  \\
\vspace*{3mm}
        THE $\lambda\Phi^4$ PHASE TRANSITION}

\author{M. Consoli, N. M. Stivala and D. Zappal\'a}

\address{ Dipartimento di Fisica - Universita' di Catania; \\
 Istituto Nazionale di Fisica Nucleare - Sezione di Catania;\\
Corso Italia, 57 - I 95129 Catania - Italy}
\newcommand{\beq}{\begin{equation}}
\newcommand{\eeq}{\end{equation}}

\maketitle

\begin{abstract}
We reanalyze the problem of the cosmological constant associated with the
phase transition in
a self-interacting scalar theory. It is
pointed out that
 the generally accepted ``triviality'' of $(\lambda\Phi^4)_4$ implies a
first-order
phase transition. As a consequence, Spontaneous Symmetry
Breaking can be consistent with zero
cosmological constant if one assumes that it vanishes in the symmetric phase
$\langle\Phi\rangle=0$.

\end{abstract}
\vfill\eject

\widetext

It is generally believed that Spontaneous Symmetry Breaking
(SSB) induces a huge cosmological constant
$\Delta$
in the Einstein equations, in strong
contradiction with the experimental result \cite{weinberg}. More precisely,
by denoting the vacuum field $\phi=\langle\Phi\rangle$
(for a one-component $\lambda\Phi^4$ theory)
 one expects a cosmological constant in the
physical broken phase at $\phi=v$
\beq
    \Delta(v)=\Delta(0)+ 8\pi G_N W
\eeq
where $G_N$ is the Newton constant and
\beq
         W=V_{\rm eff}(v)-V_{\rm eff}(0)
\eeq
represents the difference in the vacuum
energy densities of the broken and symmetric phases as
deduced from the effective potential
$V_{\rm eff}(\phi)$.
As discussed by Weinberg \cite{weinberg},the problem of the cosmological
constant is usually considered in
connection with the additional assumption
\beq
       \Delta(0)=0
\eeq
In this case, in the framework of a second order phase-transition based on
the classical potential ($\sigma<0$, $\lambda_o>0$ and
${{1}\over{2}}\rho^2=\phi^+\phi$ for a complex isodoublet scalar field)
\beq
 V_{\rm cl}(\rho)={{1}\over{2}}\sigma\rho^2+{{\lambda_o}\over{4}}\rho^4
\eeq
one finds ($v^2=-{{\sigma}\over{\lambda_o}}$)
\beq
               W =-{{\lambda_ov^4}\over{4}} <0
\eeq
or, by defining the Higgs mass from
\beq
           m^2_h=2\lambda_ov^2
\eeq
the final form
\beq
               W =-{{m^2_h v^2}\over{8}}
\eeq
In the presence of perturbative quantum corrections
Eq.(7) is still valid provided
$m_h$ and $v$ are identified with
the physical Higgs mass and the physical vacuum field
$v\sim\sqrt { {{1}\over{G_F\sqrt 2}}}\sim 246.2$ GeV. In this case,
 by taking
into account the
present lower limit from the direct Higgs search at LEP ($m_h>64.5$ GeV
\cite{glasgow}) one finds $|W|> 3\cdot 10^7~GeV^4$, about 54 orders of
magnitude larger than the experimental bound \cite{weinberg}
$|W|<10^{-47}~GeV^4$. The strong contradiction with the experimental
result shows that at least
one of the theoretical assumptions in Eqs.(1-7) should be
abandoned. To this end, one may consider different alternatives as, for
instance:
\par~~~a) to change gravity, modifying the relation
between $\Delta$ and the vacuum energy in Eq.(1)
\par~~~b) to relax the
condition $\Delta(0)=0$ in order to reconcile a large $|W|$ with
 a very small $\Delta(v)$
\par~~~c) to change the expression for the difference of vacuum
 energies in Eq.(7)

Without considering here the alternative a) (to this end, see \cite{weinberg}
sect.VII) we shall first concentrate on b). To understand the implications
 of a non-zero $\Delta(0)$, quite unrelated to the phenomenon of SSB, let us
consider the case of a free
scalar field with mass $m$
discussed in sect.I of
\cite{weinberg} where
one finds an unsubtracted vacuum energy
\beq
E_o={{1}\over{2}}\int~{{d^3k}\over{(2\pi)^3}}\sqrt{k^2+m^2}=
{{\Lambda^4}\over{16\pi^2}} +O(m^2\Lambda^2,m^4)
\eeq
in terms of the ultraviolet cutoff $\Lambda$ which truncates the sum of the
zero-point energies of all normal modes. Thus we get
\beq
        \Delta(0)=\Delta_o+ 8\pi G_N E_o
\eeq
where we have indicated with $\Delta_o$ the unobservable value of the bare
cosmological constant. As discussed in \cite{weinberg}, the assumption that
general relativity remains valid up to the Planck scale suggests the order
of magnitude extimate
\beq
            \Lambda \sim {{1}\over{ \sqrt {8\pi G_N} }}
\eeq
leading to
\beq
    E_o= 2^{-10}\pi^{-4}G^{-2}_N\sim 2\cdot 10^{71}GeV^4
\eeq
In the case of $\lambda\Phi^4$ theory
Eq.(9) is replaced by
\beq
        \Delta(0)=\Delta_o+ 8\pi G_N V_{\rm eff}(0)
\eeq
and we expect
$V_{\rm eff}(0)\sim E_o$, i.e. close to the free field case,
at least for weak bare coupling, since the vacuum energy is free of infrared
divergences.
{}From the previous discussion it is clear that
the assumption of a non-zero $\Delta(0)$ {\it worsens}
the problem. Indeed, its typical size is
about 64 orders of magnitude larger than
the contribution of SSB to the physical value of
$\Delta(v)$ in Eq.(1).
Therefore, we are forced to consider the same scenario considered by Veltman
\cite{veltman}
where, for some unknown reason, $\Delta_o$ cancels
exactly (better than 64 decimals) the genuine zero-point contribution to the
vacuum energy and this preliminary
assumption is essential for a meaningful analysis of SSB. Most likely,
a deeper understanding of the underlying physics
will impose the introduction
of some additional symmetry or the use of a
regularization scheme (such as dimensional regularization) where quartic and
quadratic divergences are absent. Here, we shall only limit
ourselves to adopt the simple recipe of assuming a vanishingly small
$\Delta(0)=0$ and concentrate on the implications for SSB of
\beq
              |W|< 10^{-47}GeV^4
\eeq
which, whenever accepted,
 requires a drastic modification of Eq.(7). In particular, we shall
explore the possibility that
\beq
         W=V_{\rm eff}(v)-V_{\rm eff}(0)=0
\eeq
which is forbidden within the framework of a second order phase transition.

To conveniently discuss the problem of the phase transition
in $\lambda\Phi^4$ theories
let us recall some rigorous results for the (one-component) lattice field
theory
described by the Euclidean action
\beq
   S =a^4 \sum_x [{{1}\over{2a^2}}\sum_{\mu}(\Phi(x+a\hat e_{\mu}) - \Phi(x))^2
+
{{\sigma}\over{2}}\Phi^2(x)  +
{{\lambda_o}\over{4}} \Phi^4(x) - J \Phi(x) ]
\eeq
where $x$ stands for a general lattice site and $a$ denotes the lattice
spacing. To study SSB, the basic quantity to
compute is the VEV of the bare scalar field $\Phi(x)$ (B=bare)
\beq
      \langle \Phi \rangle _J = \phi_B(J)
\eeq
in the presence of an external source whose strength $J(x)=J$ in Eq.(15)
is $x$-independent.
Determining $\phi_B(J)$
at several $J$-values is equivalent \cite{call2,huang} to inverting the
relation
\beq
        J=J(\phi_B)={{dV_{\rm eff}}\over{d\phi_B}}
\eeq
involving the effective potential $V_{\rm eff}(\phi_B)$. Starting from
the action in Eq.(15), the effective potential of the theory
can be {\it rigorously} defined for the lattice theory, up to an arbitrary
integration constant.
 In fact, the above procedure is precisely equivalent
to compute the Legendre transform of the generating functional
for connected
Green's function which has the meaning
(up to a space-time factor) of the
energy density in the presence of a constant source $J$.
In this framework, the occurrence of SSB is determined
by exploring the properties of the function
\beq
               \phi_B(J)=-\phi_B(-J)
\eeq
in connection with its limiting behaviour at zero external source
\beq
  \lim_{J\to 0^{\pm}}~\phi_B(J)=\pm v_B \neq 0
\eeq
over a suitable range of the bare parameters
$\sigma,\lambda_o$ appearing in the lattice action Eq.(15).

In the symmetric phase $\langle\Phi\rangle=0$,
the existence of the $(\lambda\Phi^4)_4$ {\it critical point} can be
stated for the lattice theory
on the basis of rigorous formal arguments (see chapt.17 of \cite{glimm}).
Namely,
for any $\lambda_o > 0$, a critical value $\sigma_c=\sigma_c(\lambda_o)$
exists,
such that the quantity $m(\sigma,\lambda_o)$ defined as
\beq
m(\sigma,\lambda_o))=
-\lim_{|x-y|\to\infty}~{ {\ln\langle\Phi(x)\Phi(y)\rangle} \over{|x-y|} }
\eeq
is a continuous, monotonically decreasing, non-negative
 function of $\sigma$, for $\sigma$ approaching $\sigma_c$ from above
and one
has $m(\sigma,\lambda_o)=0$ for $\sigma= \sigma_c(\lambda_o)$.
For $\sigma > \sigma_c(\lambda_o)$,
$m(\sigma,\lambda_o)$ is the gap in the energy spectrum
in the
symmetric phase (where $\langle\Phi\rangle=0$).

The basic problem
of the phase transition  in the {\it cutoff} theory
concerns the relation between $\sigma_c(\lambda_o)$ and the value
marking the onset of SSB, say $\sigma_s(\lambda_o)$,
defined as the supremum of the values of $\sigma$ at which Eq.(19)
possesses non-vanishing solutions for a given $\lambda_o$.
It should be obvious that, in general, $\sigma_s$ and $\sigma_c$ correspond to
basically
different quantities. Indeed, $\sigma_c$ defines the limiting situation
where there is no gap for the first excited state in the symmetric phase
$\langle\Phi\rangle=0$ whereas
$\sigma_s$ can only be determined after computing
 the relative magnitude of the energy density in the
symmetric and non-symmetric vacua.
For $\sigma=\sigma_c(\lambda_o)+\epsilon~$ is the symmetric phase stable
for any $\epsilon > 0$ ?
A widely accepted point of view, based on the picture of
a second-order Ginzburg-Landau
phase transition with perturbative quantum corrections,
 is that, indeed, the
system is in the broken phase at $\sigma=\sigma_c-\mu^2$ for any $\mu^2>0$,
thus implying
 $\sigma_s=\sigma_c$. In this case, $-\mu^2$ represents the ``negative
mass squared'' frequently used to describe SSB in the $\lambda\Phi^4$ theory,
related to the second derivative of the effective potential at the origin
in field space $\phi_B=0$.

An alternative possibility is the following. In general, the effective
potential is not a finite order polynomial function of the vacuum field
$\phi_B$ and (see chapt.1 of \cite{book} and chapt.4 of \cite{toledo} )
 one should consider the
more general structure
\beq
V_{\rm eff}(\phi_B)-V_{\rm eff}(0)
= {{1}\over{2}}a\phi^2_B +{{1}\over{4}}b\phi^4_B
+{{1}\over{6}}c\phi^6_B+...
\eeq
where $a,b,c,..$ depend on the bare parameters ($\sigma,\lambda_o$) so that
there are several patterns in the phase diagram. In particular,
if the coefficient $b$ can become negative, even though $a>0$, there is a first
order phase transition and the system dives in the broken phase passing
through a degenerate configuration where
$V_{\rm eff}(\pm v_B)=V_{\rm eff}(0)$. A remarkable example of this situation
was provided in \cite{halperin}. There,
 the superconducting phase transition was predicted to be (weakly) first
order, because of the effects of the intrinsic fluctuating magnetic field
which induce a negative fourth order coefficient in the free energy when the
coefficient of the quadratic term is still positive. The predicted
effect is too small to be measured but, conceptually, is extremely relevant.

Thus, on general grounds, one may consider the possibility that
$\sigma_c<\sigma_s$ even though
 for $\sigma > \sigma_c$ the symmetric phase of the
lattice $(\lambda\Phi^4)_4$ has still a mass gap $m(\sigma)> 0$ and an
exponential decay of the two-point correlation function.
The subtlety is that the general theorem 16.1.1 of
ref.\cite{glimm}, concerning the possibility of deducing the uniqueness of
the vacuum in the presence of a non-vanishing mass gap in the symmetric phase,
holds for the {\it continuum} theory.
Namely, by introducing the variable
$t=\ln{{a_o}\over{a}}$, $a_o$ being a fixed length scale, and defining the
continuum limit as a suitable path in the space of the
bare parameters $\sigma=\sigma(t)$, $\lambda_o=\lambda_o(t)$,
the mass gap in the symmetric phase has to vanish
i.e.
\beq
\lim_{t\to \infty} ~m(\sigma(t),\lambda_o(t))=0
\eeq
to consistently account for SSB in quantum field theory.
However, at any finite value of the ultraviolet cutoff
$\Lambda\sim 1/a$ there is no reason why $m$
should vanish {\it before} the system being in the broken phase thus
opening the possibility that $\sigma_c<\sigma_s$.
Actually, in a cutoff theory,
the zero-mass limit in Eq.(22) can be replaced by the more general condition
\beq
\lim_{t\to \infty} ~{ {m(\sigma(t),\lambda_o(t))}\over
{m_h(\sigma(t),\lambda_o(t))}}=0
\eeq
since the physical scale of the continuum theory is determined by the limiting
value of the Higgs mass $m_h$ and thus SSB requires $m$, the mass gap in the
symmetric phase, to become infinitesimal in units of $m_h$, the mass gap in
the broken phase (in Renormalization Group (RG) language this means that
$m$ and $m_h$ do not belong to the same {\it universality class} in the limit
$t\to \infty$).

Summarizing, on the basis of the previous discussion, it is by no means obvious
that the phase transition in $(\lambda\Phi^4)_4$ is of second-order.
Rather, an explicit calculation of the energy density in the symmetric and
broken phase is required. As shown in \cite{zeit,zappa}, by evaluating the
effective potential in those approximations consistent with the generally
accepted ``triviality''  of the theory (one-loop potential, gaussian
approximation, post-gaussian calculations \cite{rit2}, where the
Higgs propagator $G(x,y)$ is properly optimized
at each value of $\phi_B=\langle\Phi\rangle$, by solving the corresponding
non-perturbative gap-equation $G=G_o(\phi_B)$) one finds that the massless
theory at $\sigma=\sigma_c$ lies well within the broken phase and
that the phase transition is actually first-order, in contrast with the
perturbative indications. This conclusion and the basic inadequacy of the
perturbative approach are also confirmed by precise
lattice computations of the slope of the effective potential \cite{andro,cea}
near the critical line.

By using the results of \cite{zeit} one finds that the effective
potential in terms of the `` renormalized'' vacuum field $\phi_R$ defined
through
\beq
{{d^2V_{\rm eff}}\over{d\phi^2_R}}|_{\phi_R=\pm v_R}=m^2_h
\eeq
is given by the simple expression \cite{note2}
\beq
V_{\rm eff}(\phi_R)-V_{\rm eff}(0)
=\pi^2\zeta^2\phi^4_R(\ln{{\phi^2_R}\over{v^2_R}}-
{{1}\over{2}})+{{1}\over{4}}(\zeta-1)m^2_h\phi^2_R
(1-{{\phi^2_R}\over{2v^2_R}})
\eeq
$\zeta$ being defined through
\beq
                 m^2_h=8\pi^2\zeta v^2_R
\eeq
For all positive values of $\zeta$ the values $\phi_R=\pm v_R$ are minima
of the effective potential. The minimum has a lower energy than the origin
$\phi_R=0$ if $\zeta<2$. At $\zeta=2$ (corresponding to the value of the
bare mass $\sigma=\sigma_s$ discussed above)
there is a phase transition to the
broken symmetry phase and at $\zeta=1$ (corresponding to $\sigma=\sigma_c$)
one finds the Coleman-Weinberg
regime \cite{CW} where there is no intrinsic scale in the
symmetric phase. The range $0<\zeta<1$
(corresponding to values $\sigma<\sigma_c$
for which the theory cannot be quantized in its symmetric phase) is allowed
by the RG analysis of the effective potential and cannot
be discarded. Finally,
the ground state energy at the minimum can be conveniently expressed as
\beq
 W=V_{\rm eff}(v_R)-V_{\rm eff}(0)=-{{m^2_hv^2_R}\over{16}}(2-\zeta)
\eeq
{}From Eq.(27), we see that in
the limit $\zeta \to 0^+$ ( the ``extreme double well''
limit) we reobtain Eq.(7). However, differently from Eq.(7),
Eq.(27) can be consistent with $W=0$ just at the phase
transition $\zeta=2$.

In conclusion,
the structure of the effective potential of
$(\lambda\Phi^4)_4$ (in those
approximations consistent with ``triviality'') provides the indication for
a first-order phase transition so that SSB can coexist with a vanishing
vacuum energy. In this case,
by accepting the same point of view of many authors,
namely $\Delta(0)=0$, one can explain the observed result $\Delta(v_R)=0$ if
our physical vacuum corresponds just to the phase transition $\zeta=2$.

\vfill
\eject

\end{document}